\documentclass[
  pre,
  a4paper,
  twocolumn,
  showpacs,
  amsmath
]{revtex4}

\usepackage{times}
\usepackage{graphicx}

\begin{document}

\title{Hydrodynamic coupling of two rotating spheres trapped in harmonic
potentials}

\author{Michael Reichert}
\email{michael.reichert@uni-konstanz.de}
\affiliation{Fachbereich Physik, Universit\"at Konstanz, 
  D-78457 Konstanz, Germany}
\author{Holger Stark}
\affiliation{Fachbereich Physik, Universit\"at Konstanz, 
  D-78457 Konstanz, Germany}

\date{\today}

\begin{abstract}
We theoretically study in detail the hydrodynamic coupling of two 
equal-sized colloidal spheres at low Reynolds numbers assuming the 
particles to be harmonically trapped with respect to both their positions 
and orientations. By taking into account the rotational motion, we
obtain a rich spectrum of collective eigenmodes whose properties we 
determine on the basis of pure symmetry arguments. Extending recent 
investigations on translational correlations
[J.-C.~Meiners and S.~R.~Quake, Phys.~Rev.~Lett.~{\bf 82}, 2211 (1999)],
we derive the complete set of auto- and cross-correlation functions 
emphasizing the coupling of rotation to translation
which we illustrate in a few examples. An important feature of our 
system is the self-coupling of translation and rotation of one
particle mediated by the neighboring particle that is clearly visible
in the appropriate auto-correlation function. This coupling is
a higher-order effect and therefore not included in the widely used
Rotne-Prager approximation for the hydrodynamic mobilities.
\end{abstract}


\maketitle

\section{Introduction}

Colloids are widely used to model atomic systems
\cite{pus91,poo96}. However, there is one feature specific to colloidal
suspensions which distinguishes them fundamentally from atomic
systems: the so-called hydrodynamic interactions
\cite{dho96,hap73}. A particle moving in a viscous fluid creates a
long-range flow field around itself
through which it interacts with other particles. Thus,
hydrodynamic interactions constitute a complicated many-body problem
since the motion of one particle depends on the translations and
rotations of all the other particles in the fluid \cite{fel88}.

The central quantities in the description of hydrodynamic interactions
are the mobility or friction tensors which connect in a linear
response scheme all the forces and torques acting on the particles to
their linear and angular velocities \cite{bre63+64}. In the present
paper, we draw special attention to the rotational degree of freedom
and how it couples to translation. Its physical consequences have
rarely been treated in literature \cite{jon88+89, deg95,koen01}.

In many physically interesting systems, hydrodynamic interactions play
an important role. Whenever dynamic effects in suspensions are
studied, hydrodynamic interactions have to be taken into account. 
Examples are sedimenting particles \cite{lad93,bre99} or the
apparent attractive interaction between like-charged
spherical particles mediated by a single, like-charged wall
\cite{squi00}.

In recent years, a new method called microrheology has been
developed \cite{mas97} and applied to several biological systems
\cite{git97,schmi00}.
It is used to determine rheological properties of viscous
and viscoelastic media by tracking the trajectory of embedded probe
particles and calculating the time-dependent position
correlations. The so-called two-point microrheology
employs a system of two colloidal particles, which has several
advantages over using only a single particle; in particular, the
correlated motions of two tracer particles reflect the bulk rheology
of the medium they are embedded in more accurately
\cite{cro00,lev00}. Many experiments based on this technique have been 
carried out to study soft media \cite{mei99,bar01,hen01,star03}.
Furthermore, this method was used to measure the corrections to the
diffusion coefficients in a system of two spheres due to hydrodynamic
interactions \cite{cro97} and to study the hydrodynamic coupling of
two spheres to a wall \cite{duf00}. 

\begin{figure}[b]
\centering
\includegraphics[width=\columnwidth]{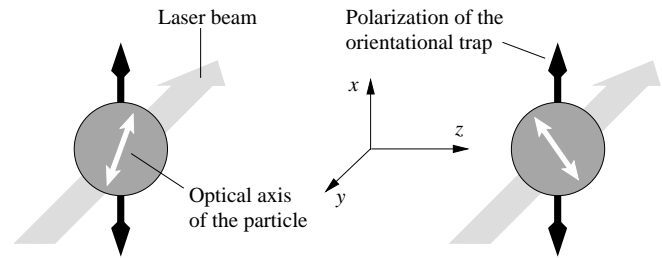}
\caption{Sketch of a setup to monitor rotational correlations of two 
birefringent particles. The particles are situated on the $z$-axis. 
The laser beam of the orientational optical trap is directed along the
$y$-axis and its polarization points into $x$-direction.
To monitor transversal correlations, the direction of observation is
along the $y$-axis. Pure longitudinal rotational correlations are only
observable along the $z$-axis.}
\label{fig:setup}
\end{figure}

The physical systems we have in mind are colloidal suspensions of 
birefringent spherical particles. E.g., they are realized by
polymerizing the nematic order in liquid-crystal droplets
\cite{cai01,mer02}. In contrast to conventional
isotropic colloids, both their positions and orientations can be
manipulated by optical tweezers. In particular, the orientation is
controlled with optical traps generated by linearly polarized laser
light and is detected, e.g., with crossed polarizers \cite{cai01,mer02,juo99}
(see Fig.~\ref{fig:setup}).

In the present paper, we consider a system of two trapped spheres of
equal size. We give a complete analytic description of the coupled
translational and rotational dynamics. Due to the axial symmetry, the
longitudinal motions parallel to the particle-particle axis decouple
completely from transversal motions. In
addition, there is no coupling of translation and rotation in
longitudinal motions. This greatly reduces the complexity of the
problem. We determine the complete set of collective eigenmodes of the
system and calculate the correlations for particles undergoing thermal
fluctuations. Our work extends recent investigations on correlations
in translational fluctuations \cite{mei99,bar01,hen01}. By taking into
account the rotational degrees of freedom, we obtain a rich spectrum
of collective modes and correlation functions.

The outline of this paper is as follows. In
Sec.~\ref{sec:hydro_int}, we first summarize the basic equations for
the description of hydrodynamic interactions in general. Then,
we study in detail the two-particle system and show by symmetry
arguments how it decouples into the aforementioned 
longitudinal and transversal motions. 
Assuming that both the particles' positions and orientations are
trapped in harmonic potentials, we determine and discuss their
collective eigenmodes in Sec.~\ref{sec:eigenmodes}. Finally, in 
Sec.~\ref{sec:correl}, we use Langevin dynamics to calculate
correlation functions for the thermally fluctuating positions and
orientations of the trapped particles and point out their interesting
features.

\section{Hydrodynamic interactions}
\label{sec:hydro_int}

\subsection{Fundamental equations}
\label{subsec:fund_eq}

In the regime of low Reynolds numbers and on the Brownian time scale,
the flow of an incompressible fluid with viscosity $\eta$ obeys the
Stokes or creeping flow equations \cite{dho96},
\begin{equation}
\label{eq:stokes}
\eta\nabla^{2}\vec{u}-\nabla p = \vec{0} \, , \quad
\nabla\cdot\vec{u} = 0 \, ,
\end{equation}
where $\vec{u}$ is the flow field and $p$ the hydrodynamic pressure.
Stokesian dynamics describes overdamped motion in a viscous fluid.
In the following, we consider motions of particles in an unbounded and
otherwise quiescent fluid, i.e., $\vec{u}=\vec{0}$ at infinity.

Imposing stick boundary conditions on the surfaces of all $N$
particles suspended in the fluid at positions $\vec{r}_{i}$
($i=1\ldots,N$), the motions of the particles are mutually coupled via the
flow field. Due to the linearity of Eqs.~(\ref{eq:stokes}), the
translational and rotational velocities of the particles,
$\vec{v}_{i}$ and $\vec{\omega}_{i}$, depend linearly on all external
forces and torques acting on the particles,
$\vec{F}_{j}$ and $\vec{T}_{j}$ \cite{bre63+64}:
\begin{subequations}
\label{eq:(v,omega)=(mutt,mutr,murt,murr)*(F,T)}
\begin{align}
\vec{v}_{i} & = \sum\limits_{j=1}^{N}\left(\mutt_{ij}\vec{F}_{j}
+\mutr_{ij}\vec{T}_{j}\right) ,
\\
\vec{\omega}_{i} & = \sum\limits_{j=1}^{N}\left(\murt_{ij}\vec{F}_{j}
+\murr_{ij}\vec{T}_{j}\right) .
\end{align}
\end{subequations}
The central quantities constituting the mutual coupling of translation
and rotation (denoted by superscripts $\text{t}$ and $\text{r}$) of
two particles $i$ and $j$ are the $3\times 3$ mobility tensors
$\mutt_{ij}$, $\murr_{ij}$, $\mutr_{ij}$, and $\murt_{ij}$. They
depend on the current spatial configuration of all particles, i.e.,
the set of position vectors $\{\vec{r}_{1},\ldots,\vec{r}_{N}\}$ in
the case of spherical particles. 

To introduce a more compact notation, we define the $6N$-dimensional
vectors $\Vec{V}=[\vec{v}_{1},\ldots,\vec{v}_{N},\vec{\omega}_{1},
\ldots,\vec{\omega}_{N}]$ and $\Vec{F}=[\vec{F}_{1},\ldots,
\vec{F}_{N},\vec{T}_{1},\ldots,\vec{T}_{N}]$. Then,
Eqs.~(\ref{eq:(v,omega)=(mutt,mutr,murt,murr)*(F,T)}) take the form
\begin{equation}
\label{eq:V=M*F}
\Vec{V}=\Mat{M}\Vec{F}
\end{equation}
with the $6N\times 6N$ mobility matrix
\begin{equation}
\Mat{M}=
\left[
\begin{array}{cc}
[\mutt_{ij}] & [\mutr_{ij}] \\[1ex]
[\murt_{ij}] & [\murr_{ij}]
\end{array}
\right]
\, ,
\end{equation}
where the four blocks of $\Mat{M}$ consist each of $N\times N$ matrices
whose elements are the $3\times 3$ mobility tensors; e.g., 
\begin{equation}
[\mutt_{ij}]=
\left[
\begin{array}{ccc}
\mutt_{11} & \hdots & \mutt_{1N} \\
\vdots & \ddots & \vdots \\
\mutt_{N1} & \hdots & \mutt_{NN}
\end{array}
\right]
\, .
\end{equation}
According to the reciprocal theorem of Lorentz, the mobility tensors
fulfill the symmetry relations \cite{bre63+64}
\begin{equation}
(\mutt_{ij})\T = \mutt_{ji} \, , \quad
(\murr_{ij})\T = \murr_{ji} \, , \quad
(\mutr_{ij})\T = \murt_{ji} \, ,
\label{eq:mobil-tensors_symm}
\end{equation}
where $(\mutt_{ij})\T$ denotes the transpose of $\mutt_{ij}$,
etc. Thus, the entire $6N\times 6N$ matrix $\Mat{M}$ is symmetric.

In the overdamped limit discussed here, the work done on the particles
by the external forces and torques $\Vec{F}$ is completely dissipated
in the fluid, so the energy dissipation rate
$\sum_{i}(\vec{v}_{i}\cdot\vec{F}_{i}+\vec{\omega}_{i}\cdot\vec{T}_{i})
=\Vec{V}\cdot\Vec{F}=\Vec{V}\cdot\Mat{M}^{-1}\Vec{V}$ has to be
positive. 
Therefore, the friction matrix $\Mat{M}^{-1}$ and hence the mobility
matrix $\Mat{M}$ itself are positive definite.

There a several methods to calculate the mobility tensors for a given
many-particle system, e.g., the concept of reflected flow fields
using the gradient expansion technique \cite{dho96,bre63+64} or the
method of induced force multipoles \cite{fel88,cich94}. The latter was
implemented in the numerical library {\sc hydrolib} \cite{hin95} which
calculates the mobility or friction matrix for a given configuration
of spheres.

\subsection{Two-sphere system}
\label{subsec:two_spheres}

In the following, we consider a system of two equal-sized spheres. Let
$\vec{r}$ be the vector connecting the centers of the two spheres,
pointing from sphere~1 to sphere~2, and $r$ the center-to-center
distance. Then, $\hatvecr=\vec{r}/r$ is the unit vector along the line
of centers. Due to the rotational symmetry about the axis $\hatvecr$
and the different parities of polar ($\vec{v}_{i}$ and $\vec{F}_{i}$) and 
pseudo-vectors ($\vec{\omega}_{i}$ and $\vec{T}_{i}$), the mobility
tensors can be written as \cite{jef84}
\begin{subequations}
\label{eq:mobil-tensors_form}
\begin{align}
\mutt_{ij}(\vec{r}) & = \muttpara_{ij}(r)\,\hatvecr\otimes\hatvecr
+\muttperp_{ij}(r)\,(\mat{1}-\hatvecr\otimes\hatvecr) \, , \\
\murr_{ij}(\vec{r}) & = \murrpara_{ij}(r)\,\hatvecr\otimes\hatvecr
+\murrperp_{ij}(r)\,(\mat{1}-\hatvecr\otimes\hatvecr) \, , \\
\mutr_{ij}(\vec{r}) & = \mutrperp_{ij}(r)\,\hatvecr\!\times \, .
\end{align}
\end{subequations}
The mobility coefficients $\muttparaperp_{ij}$, $\murrparaperp_{ij}$, and
$\mutrperp_{ij}$ ($i,j=1,2$) are scalar
functions depending only on the center-to-center distance $r$. They
describe motions parallel and perpendicular to the axis, respectively.
Using the general symmetry relations (\ref{eq:mobil-tensors_symm}) and the 
fact that particles 1 and 2 are identical, the mobility coefficients
obey \cite{jef84}
\begin{subequations}
\label{eq:mobil-coeff_invsymm}
\begin{alignat}{3}
\muttparaperp_{11} &= \muttparaperp_{22} , &\hspace{2ex}
\murrparaperp_{11} &= \murrparaperp_{22} , &\hspace{2ex}
\mutrperp_{11}     &= -\mutrperp_{22}    , \\
\muttparaperp_{12} &= \muttparaperp_{21} , &
\murrparaperp_{12} &= \murrparaperp_{21} , &
\mutrperp_{12}     &= -\mutrperp_{21}    .
\end{alignat}
\end{subequations}

\begin{figure}[t]
\centering
\includegraphics[width=\columnwidth]{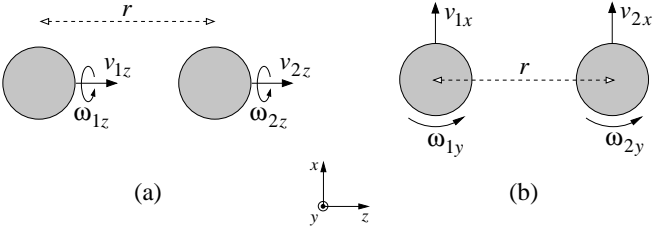}
\caption{Definition of the applied Cartesian coordinate system. The
$z$-direction is along the center-to-center line of the two spheres. 
Longitudinal motions (a) decouple from transversal motions
(b). Furthermore, there is no coupling of translation and
rotation for longitudinal motions.}
\label{fig:coord}
\end{figure}

With the explicit expressions (\ref{eq:mobil-tensors_form}), the
dynamics of the two-sphere system described by
Eq.~(\ref{eq:V=M*F}) separates into longitudinal motions parallel and
transversal motions perpendicular to the particle-particle axis
$\hat{\vec{r}}$. Furthermore, longitudinal translations and
longitudinal rotations (i.e., rotations about the axis) do not couple
to each other because of the different parities of translations (polar
vectors) and rotations (axial vectors). In a Cartesian coordinate 
system with the $z$-direction pointing along $\hatvecr$ 
(see Fig.~\ref{fig:coord}), we obtain from 
Eqs.~(\ref{eq:(v,omega)=(mutt,mutr,murt,murr)*(F,T)}), 
(\ref{eq:mobil-tensors_form}), and (\ref{eq:mobil-coeff_invsymm}) 
for longitudinal motions
\begin{equation}
\label{eq:(vz,omegaz)=(mupara)*(Fz,Tz)}
\left[
\begin{array}{c}
v_{1z}\muparablind \\[0.5ex]
v_{2z}\muparablind \\[0.5ex]
\omega_{1z}\muparablind \\[0.5ex]
\omega_{2z}\muparablind
\end{array}
\right]
\hspace{-0.4ex}
=
\hspace{-0.4ex}
\left[
\begin{array}{cccc}
\muttpara_{11} & \muttpara_{12} & 0 & 0 \\[0.5ex]
\muttpara_{12} & \muttpara_{11} & 0 & 0 \\[0.5ex]
0 & 0 & \murrpara_{11} & \murrpara_{12} \\[0.5ex]
0 & 0 & \murrpara_{12} & \murrpara_{11}
\end{array}
\right]
\hspace{-0.7ex}
\left[
\begin{array}{c}
F_{1z}\muparablind \\[0.5ex]
F_{2z}\muparablind \\[0.5ex]
T_{1z}\muparablind \\[0.5ex]
T_{2z}\muparablind
\end{array}
\right] ,
\end{equation}
where $v_{1z}$ is the $z$-component of the translational velocity
of particle~1, etc. In transversal motions, translations along the
$x$-direction are coupled to rotations about the $y$-axis:
\begin{equation}
\label{eq:(vx,omegay)=(muperp)*(Fx,Ty)}
\left[
\begin{array}{c}
v_{1x}\muperpblind \\[0.5ex]
v_{2x}\muperpblind \\[0.5ex]
\omega_{1y}\muperpblind \\[0.5ex]
\omega_{2y}\muperpblind
\end{array}
\right]
\hspace{-0.4ex}
=
\hspace{-0.4ex}
\left[
\begin{array}{rrrr}
 \muttperp_{11} & \muttperp_{12} & -\mutrperp_{11} & -\mutrperp_{12} \\[0.5ex]
 \muttperp_{12} & \muttperp_{11} &  \mutrperp_{12} &  \mutrperp_{11} \\[0.5ex]
-\mutrperp_{11} & \mutrperp_{12} &  \murrperp_{11} &  \murrperp_{12} \\[0.5ex]
-\mutrperp_{12} & \mutrperp_{11} &  \murrperp_{12} &  \murrperp_{11}
\end{array}
\right]
\hspace{-0.7ex}
\left[
\begin{array}{c}
F_{1x}\muperpblind \\[0.5ex]
F_{2x}\muperpblind \\[0.5ex]
T_{1y}\muperpblind \\[0.5ex]
T_{2y}\muperpblind
\end{array}
\right] .
\end{equation}
Translations along the $y$-direction and rotations about the $x$-axis
obey an equivalent system of equations.

Defining a four-dimensional velocity vector $\Vec{v}$ and a force
vector $\Vec{f}$, we abbreviate
Eqs.~(\ref{eq:(vz,omegaz)=(mupara)*(Fz,Tz)}) and 
(\ref{eq:(vx,omegay)=(muperp)*(Fx,Ty)}), respectively, by
\begin{equation}
\label{eq:v=m*f}
\Vec{v}=\Mat{m}\Vec{f} \, ,
\end{equation}
where the appropriate $4\times 4$ mobility matrix $\Mat{m}$ is
still symmetric. Thus, rotational symmetry reduces the full 
$12\times 12$ problem (\ref{eq:V=M*F}) to two $4\times 4$ problems,
where Eq.~(\ref{eq:(vz,omegaz)=(mupara)*(Fz,Tz)}) is essentially a
$2\times 2$ problem. 

Accordingly, the energy dissipation rate
$\Vec{V}\cdot\Mat{M}^{-1}\Vec{V}$ splits up into a sum of three
independent terms of the form $\Vec{v}\cdot\Mat{m}^{-1}\Vec{v}$, so
the single mobility matrices $\Mat{m}$ have to be positive definite.

\subsection{Mobilities in Rotne-Prager approximation}
\label{subsec:rotne-prager}

For a system of two particles, the mobility tensors can be
calculated, e.g., with the method of reflections based on the
Fax\'{e}n theorem. It provides a systematic expansion of the mobility
tensors in powers of the inverse particle separation (for a detailed
description, see, e.g., Ref.~\cite{dho96}).

The leading order in the far-field approximation of the mobilities is
the well-known Oseen tensor which considers the particles as
point-like and hence does not include rotations. The next-higher order
is the so-called Rotne-Prager approximation. It corresponds 
to one reflection of the flow field and is therefore exact 
up to the order of $1/\rho^{3}$, where $\rho=r/a$ is the dimensionless
particle distance ($a$ is the radius of the spheres). The relevant
mobilities including the rotational degrees of freedom are presented, e.g., 
in Refs.~\cite{maz82,jef84}. They also follow straightforwardly by
extending the calculations in Ref.~\cite{dho96}, where the torques are
explicitly set to zero.

In the Rotne-Prager approximation, the self-mobilities are
identical to the Stokes coefficients for single spheres, 
\begin{align}
\label{eq:RP-muttii}
\mutt_{ii} &= \mut\mat{1} \, , \quad \timestext{where} \;\;
\mut=(6\pi\eta a)^{-1} \, , \\
\label{eq:RP-murrii}
\murr_{ii} &= \mur\mat{1} \, , \quad \timestext{where} \;\;
\mur=(8\pi\eta a^{3})^{-1} \, ,
\end{align}
and there is no self-coupling of translation and rotation, i.e.,
\begin{equation}
\label{eq:RP-mutrii}
\mutr_{ii}=\murt_{ii}=\mat{0} \, .
\end{equation}
Note that for an isolated sphere within the linear Stokes regime,
translation and rotation are not coupled; the so-called Magnus effect 
only enters via the nonlinear term in the Navier-Stokes equation
\cite{hes68}. On the other hand, in the two-sphere system, the
self-coupling exists since it is mediated by the 
second particle. However, this is an effect of higher order than
$1/\rho^{3}$ and therefore not included in the Rotne-Prager
approximation, as stated in Eq.~(\ref{eq:RP-mutrii}).

For completeness, we also give the essential cross-mobilities:
\begin{align}
\label{eq:RP-muttij}
\mutt_{12} &= \mut\bigg[\frac{3}{4\rho}\,
(\mat{1}+\hatvecr\otimes\hatvecr) +\frac{1}{2\rho^{3}}\,
(\mat{1}-3\,\hatvecr\otimes\hatvecr)\bigg] \, ,
\\[1ex]
\label{eq:RP-murrij}
\murr_{12} &= \mur\left[-\frac{1}{2\rho^{3}}\,
(\mat{1}-3\,\hatvecr\otimes\hatvecr)\right] \, ,
\\[1ex]
\label{eq:RP-mutrij}
\mutr_{12} &= \mur\left[a\,\frac{1}{\rho^{2}}\,\hatvecr\times\,\right] \, .
\end{align}

\section{Eigenmodes of two trapped spheres}
\label{sec:eigenmodes}

We aim to study thermal motions of particles that are trapped 
with respect to both their positions and orientations.
Assuming that the spatial and angular displacements are small, we
consider harmonic trap forces and torques, i.e.,
\begin{align}
\label{eq:Fialpha}
F_{i\alpha} &= -\kt r_{i\alpha} \, , \\
\label{eq:Tialpha}
T_{i\alpha} &= -\kr\chi_{i\alpha} \, ,
\end{align}
where $\kt$ and $\kr$ are positive force and torque constants ($i=1,2$ is
the particle number and $\alpha=x,y,z$ the coordinate index). 
The particle velocities are 
\begin{equation}
\label{eq:vialpha}
v_{i\alpha}=\dot{r}_{i\alpha} \, ,
\end{equation}
where the spatial coordinate $r_{i\alpha}$ denotes the displacement of
particle $i$ along direction $\alpha$.

\begin{figure}[t]
\centering
\includegraphics[width=0.9\columnwidth]{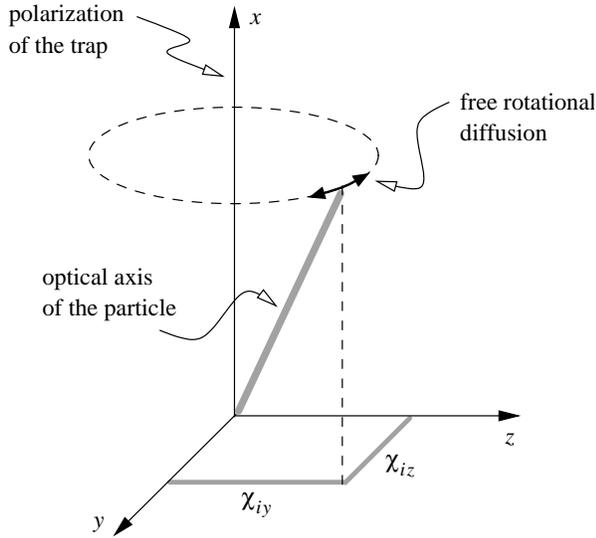}
\caption{While the rotation of the optical axis of the trapped particle
about the direction of the trap polarization 
(here pointing along the $x$-axis) is free, the deviation from the 
$x$-direction is restricted, and therefore the angular displacements 
$\chi_{iy}$ and $\chi_{iz}$ are small (see text).}
\label{fig:angles}
\end{figure}

The angle $\chi_{i\alpha}$ describes the rotation of particle $i$
about the coordinate axis $\alpha$. As long as this angular
displacement is small ($\langle\chi_{i\alpha}^{2}\rangle\ll 1$),
we can relate it to the angular velocity via
\begin{equation}
\label{eq:omegaialpha}
\omega_{i\alpha}=\dot{\chi}_{i\alpha} \, .
\end{equation}
This is a nontrivial statement and requires some explanation. The
optical axis of the trapped birefringent particle aligns along the
polarization of the laser trap pointing along, e.g., the $x$-axis
(see Fig.~\ref{fig:angles}). Any deviation of the particle orientation from
the $x$-direction is described by the angles $\chi_{iy}$ and
$\chi_{iz}$, and relaxes in the optical trap on the time scale
$(\kr\mur)^{-1}$. However, the free rotational diffusion of
the particle axis about the $x$-direction also changes the angular
displacements $\chi_{iy}$ and $\chi_{iz}$ (see Fig.~\ref{fig:angles}),
and therefore, relation (\ref{eq:omegaialpha}) does not hold in
general. Nevertheless, in the limit  considered here the orientational
relaxation is a much faster process than the free diffusion, so on the
relaxation time scale $(\kr\mur)^{-1}$, the free diffusion can be
neglected and  relation (\ref{eq:omegaialpha}) is applicable. To show
this, we note that the free rotational diffusion takes 
place on the time scale $(\kB T\mur)^{-1}$. Requiring that
$(\kr\mur)^{-1}/(\kB T\mur)^{-1}=\kB T/\kr\ll 1$ and using
the equipartition theorem yields 
$\langle\chi_{i\alpha}^{2}\rangle=\kB T/\kr\ll 1$, which is consistent 
with the initial assumption.

We define generalized coordinates
$\Vec{q}=[r_{1z},r_{2z},\chi_{1z},\chi_{2z}]$ for longitudinal modes
and $\Vec{q}=[r_{1x},r_{2x},\chi_{1y},\chi_{2y}]$ for trans\-versal
modes and combine the corresponding trap forces (\ref{eq:Fialpha}) and 
torques (\ref{eq:Tialpha}) into a four-dimensional vector
\begin{equation}
\label{eq:f=-k*q}
\Vec{f}=-\Mat{k}\Vec{q}
\end{equation}
with the diagonal force constant matrix
\begin{equation}
\label{eq:k}
\Mat{k}=
\left[
\begin{array}{cccc}
\ktblind \kt &   0 &   0 &   0  \\
           0 & \kt &   0 &   0  \\
           0 &   0 & \kr &   0  \\
           0 &   0 &   0 & \kr
\end{array}
\right]
\, .
\end{equation}
Then, the equation of motion (\ref{eq:v=m*f}) for both longitudinal and 
trans\-versal modes reads 
\begin{equation}
\label{eq:dotq=-mkq}
\dot{\Vec{q}}+\Mat{mk}\Vec{q}=\Vec{0} \, ,
\end{equation}
where $\Mat{m}$ is the appropriate mobility matrix given by
Eqs.~(\ref{eq:(vz,omegaz)=(mupara)*(Fz,Tz)}) and
(\ref{eq:(vx,omegay)=(muperp)*(Fx,Ty)}), respectively.
Note that for spatial displacements small compared to the
equilibrium particle distance, we can consider $\Mat{m}$ to be
constant.

The solutions of Eq.~(\ref{eq:dotq=-mkq}) are relaxational eigenmodes
$\e^{-\lambda_{n}t}\Vec{a}_{n}$ with relaxation rates $\lambda_{n}$ or
relaxation times $\lambda_{n}^{-1}$. They are determined by
the eigenvalue problem
\begin{equation}
\label{eq:evp}
\Mat{mk}\Vec{a}_{n}=\lambda_{n}\Vec{a}_{n} \quad (n=1,\ldots,4)
\end{equation}
of the nonsymmetric $4\times 4$ matrix $\Mat{mk}$ whose eigenvectors
$\Vec{a}_{n}$ are in general not perpendicular to each other. In the
next two sections, we will analyze the eigenmodes in detail.

\subsection{Longitudinal eigenmodes}
\label{subsec:long_modes}

To study the longitudinal eigenmodes, the traps have to be polarized
along the $x$- or $y$-direction so that the angular displacements
$\chi_{iz}$ are small. 
Since longitudinal translations and rotations are decoupled
(as discussed in Sec.~\ref{subsec:two_spheres}), we can immediately
write down the eigenvalues and eigenvectors of the four eigenmodes:
\begin{subequations}
\label{eq:evec/eval_long}
\begin{align}
\label{eq:evec/eval_long_12}
\lambda_{1/2} &= \kt\bigl(\muttpara_{11}\pm\muttpara_{12}\bigr) \, , &
\Vec{a}_{1/2} &= [1,\pm 1,0,0] \, , \\
\label{eq:evec/eval_long_34}
\lambda_{3/4} &= \kr\bigl(\murrpara_{11}\pm\murrpara_{12}\bigr) \, , &
\Vec{a}_{3/4} &= [0,0,1,\pm 1] \, .
\end{align}
\end{subequations}
They consist of relative ($-$) and collective ($+$) modes. Intuitive arguments
allow a comparison of the respective relaxation rates. E.g., when the 
spheres translate in opposite directions (relative
translational mode $\Vec{a}_{2}$), some fluid has to be pulled into or
squeezed out of the region between the two particles; or when they rotate
in opposite directions (relative rotational mode $\Vec{a}_{4}$), the
fluid between the spheres has to be sheared. On the other hand, when
the spheres translate or rotate collectively (modes $\Vec{a}_{1}$ or
$\Vec{a}_{3}$), the fluid surrounding the spheres is just 
``displaced'' or ``rotated'' as a whole. So the collective modes experience
less resistance and therefore relax faster than the relative modes.
This is in accordance with Eqs.~(\ref{eq:evec/eval_long}) which give
$\lambda_{1}>\lambda_{2}$ and $\lambda_{3}>\lambda_{4}$.

\subsection{Transversal eigenmodes}
\label{subsec:trans_modes}

The appropriate coordinates to treat the transversal
eigenmodes are $\Vec{q}=[r_{1x},r_{2x},\chi_{1y},\chi_{2y}]$,
and the mobility matrix $\Mat{m}$ is given by
Eq.~(\ref{eq:(vx,omegay)=(muperp)*(Fx,Ty)}). The polarization of the
traps has to be chosen along the $x$- or $z$-direction so that the
angles $\chi_{iy}$ are small.

The diagonalization of the nonsymmetric $4\times4$ matrix $\Mat{mk}$ is not
immediately obvious. However, symmetry arguments help to identify 
the eigenvectors. Since the two particles are identical and since
spatial coordinates and angles possess different parities, one
readily shows that the eigenvectors have to be of the form 
$[A,A,B,-B]$ or $[A,-A,B,B]$. This constraint simplifies the 
determination of the eigenmodes considerably. There are two relaxational 
modes with symmetric translation and antisymmetric rotation,
\begin{subequations}
\label{eq:evec/eval_trans_12}
\begin{align}
\label{eq:eval_trans_12}
\lambda_{1/2} &= \frac12\biggl[\kt\muttperp_{+}+\kr\murrperp_{-}
\\[-0.5ex]
& \hspace{5.5ex} \pm\sqrt{
\bigl(\kt\muttperp_{+}-\kr\murrperp_{-}\bigr)^{2}
+4\kt\kr\bigl(\mutrperp_{-}\bigr)^{2}
}
\,\,\biggr]
\, , \nonumber \\[2ex]
\label{eq:evec_trans_12}
\Vec{a}_{1/2} &= [A_{1/2},A_{1/2},B_{1/2},-B_{1/2}] \, ,
\intertext{where}
A_{1/2} &= - 2\kr\mutrperp_{-} \, , \\
B_{1/2} &= - \bigl(\kt\muttperp_{+}-\kr\murrperp_{-}\bigr) \\
& \hspace{2.5ex} \pm\sqrt{
\bigl(\kt\muttperp_{+}-\kr\murrperp_{-}\bigr)^{2}
+4\kt\kr\bigl(\mutrperp_{-}\bigr)^{2}
}
\,\, , \nonumber
\end{align}
\end{subequations}
and two modes with antisymmetric translation and symmetric rotation,
\begin{subequations}
\label{eq:evec/eval_trans_34}
\begin{align}
\label{eq:eval_trans_34}
\lambda_{3/4} &= \frac12\biggl[\kt\muttperp_{-}+\kr\murrperp_{+}
\\[-0.5ex]
& \hspace{5.5ex} \pm\sqrt{
\bigl(\kt\muttperp_{-}-\kr\murrperp_{+}\bigr)^{2}
+4\kt\kr\bigl(\mutrperp_{+}\bigr)^{2}
}
\,\,\biggr]
\, , \nonumber \\[2ex]
\label{eq:evec_trans_34}
\Vec{a}_{3/4} &= [A_{3/4},-A_{3/4},B_{3/4},B_{3/4}] \, ,
\intertext{where}
A_{3/4} &= - 2\kr\mutrperp_{+} \, , \\[1ex]
B_{3/4} &= - \bigl(\kt\muttperp_{-}-\kr\murrperp_{+}\bigr) \\
& \hspace{2.5ex} \pm\sqrt{
\bigl(\kt\muttperp_{-}-\kr\murrperp_{+}\bigr)^{2}
+4\kt\kr\bigl(\mutrperp_{+}\bigr)^{2}
}
\,\, . \nonumber
\end{align}
\end{subequations}
Here, we use the abbreviations
$\muttperp_{\pm}=\muttperp_{11}\pm\muttperp_{12}$, etc.

Without relying on the exact values of all the mobilities, we can already
infer the qualitative characteristics of the eigenmodes based on
general arguments.
The signs of the components $B_{1-4}$ are independent of the
mobility coefficients. It is $B_{1}>0$, $B_{2}<0$, $B_{3}>0$, and
$B_{4}<0$. The signs of $A_{1-4}$ depend on the signs of
$\mutrperp_{\pm}$. In Sec.~\ref{subsec:rotne-prager}, we already
discussed on the basis of the reflection method that the absolute
value of the cross-mobility $\mutrperp_{12}$ is larger than the
self-mobility $\mutrperp_{11}$. Furthermore, by considering the
special case $v_{1x}=-\mutrperp_{12}T_{2y}$ (for the geometry, see
Fig.~\ref{fig:coord}), we infer that $\mutrperp_{12}$ is
positive. Thus, $\mutrperp_{+}>0$ and $\mutrperp_{-}<0$, and we find
$A_{1/2}>0$ and $A_{3/4}<0$.

We note that all eigenvalues of $\Mat{mk}$ are positive. To prove this
statement, we require $\lambda_{n}>0$ in
Eqs.~(\ref{eq:eval_trans_12}) and (\ref{eq:eval_trans_34}) and find
the condition $\muttperp_{\pm}\murrperp_{\mp}>(\mutrperp_{\mp})^{2}$,
which is the same guaranteeing that the mobility matrix is
positive definite.

\begin{figure}[t]
\centering
\includegraphics[width=\columnwidth]{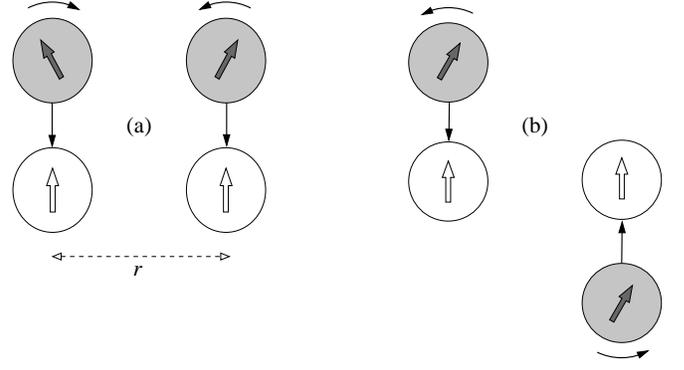}
\caption{Transversal eigenmodes of two trapped spheres (white: relaxed
equilibrium state). There are two modes with collective translation
and relative rotation (a) and two modes with relative translation and
collective rotation (b). For each mode shown, there is a complementary
mode with opposite direction of rotation.} 
\label{fig:trans_modes}
\end{figure}

\begin{figure}[b]
\centering
\includegraphics[width=\columnwidth]{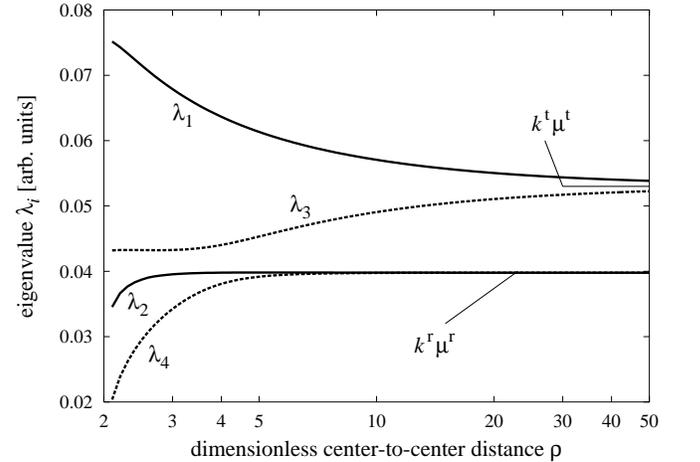}
\caption{Eigenvalues of the transversal modes
[Eqs.~(\ref{eq:eval_trans_12}) and (\ref{eq:eval_trans_34})] as a
function of the dimensionless center-to-center distance
$\rho=r/a$. The mobilities were calculated using the numerical library
{\sc hydrolib} \cite{hin95}. The trap constants are chosen as $\kt=\kr$.} 
\label{fig:eigenvalues}
\end{figure}

The eigenmodes described by Eqs.~(\ref{eq:evec_trans_12}) and
(\ref{eq:evec_trans_34}) are illustrated in
Fig.~\ref{fig:trans_modes}. The mode in Fig.~\ref{fig:trans_modes}(a)
corresponds to the eigenvector $\Vec{a}_{1}$, and the one in
Fig.~\ref{fig:trans_modes}(b) to $\Vec{a}_{3}$. Each of the modes is
complemented by a mode with opposite direction of rotation
corresponding to the eigenvectors $\Vec{a}_{2}$ and $\Vec{a}_{4}$,
respectively. Fig.~\ref{fig:trans_modes} shows the faster modes, 
i.e., $\lambda_{1}>\lambda_{2}$ and
$\lambda_{3}>\lambda_{4}$. The reason for this is simply that the
relaxations described by $\Vec{a}_{1}$ and $\Vec{a}_{3}$ show
qualitatively the same behavior as if the particles were not trapped
with respect to their orientations ($\kr=0$). Hence, $\Vec{a}_{2}$ and
$\Vec{a}_{4}$ correspond to modes where the orientational relaxation
is opposite to this ``natural'' direction of rotation. Therefore,
these modes are slower compared to their respective partners
$\Vec{a}_{1}$ and $\Vec{a}_{3}$.

These qualitative considerations are in agreement with exact numbers for
the eigenvalues which we plot in Fig.~\ref{fig:eigenvalues} as a function 
of the particle distance. Furthermore, for large distances, rotational
and transversal motions decouple since $\lambda_{1/3}$ tend towards
the single-particle translational relaxation rate $\kt\mut$ 
whereas $\lambda_{2/4}$ assume the corresponding rotational value 
$\kr\mur$. As obvious from the figure, this decoupling occurs 
for the rotational motion at shorter distances compared to translations
since rotational flow fields decay faster than translational ones.

\section{Time-correlations in Brownian motion}
\label{sec:correl}

To treat the Brownian motion of the two spheres,
time-dependent random forces and torques $\tilde{\Vec{f}}(t)$
mimicking the microscopic degrees of freedom of the surrounding
fluid have to be added. The total generalized force vector is then 
$\Vec{f}=-\Mat{k}\Vec{q}+\tilde{\Vec{f}}(t)$.
Extending Eq.~(\ref{eq:dotq=-mkq}), we obtain the Langevin-type equation 
\begin{equation}
\label{eq:ornstein-uhlenbeck}
\dot{\Vec{q}}+\Mat{mk}\Vec{q}=\Mat{m}\tilde{\Vec{f}}(t) \, ,
\end{equation}
which describes a so-called Ornstein-Uhlenbeck process \cite{ris89}.
The random force is assumed to be a Gaussian white noise and is
fully characterized by its first and second moments,
\begin{equation}
\label{eq:white_noise}
\bigl\langle\tilde{\Vec{f}}(t)\bigr\rangle = \Vec{0} \, , \quad
\bigl\langle\tilde{\Vec{f}}(t)\otimes\tilde{\Vec{f}}(t')\bigr\rangle
= 2\kB T\,\Mat{m}^{-1}\delta(t-t') \, .
\end{equation}
The fluctuation-dissipation theorem (\ref{eq:white_noise}) relates the
second moment of the fluctuating forces and torques to the friction
matrix $\Mat{m}^{-1}$ \cite{hau73}.

The formal solution of the Ornstein-Uhlenbeck process described by
Eq.~(\ref{eq:ornstein-uhlenbeck}) reads \cite{ris89}
\begin{equation}
\label{eq:q(t)}
\Vec{q}(t)=\e^{-\Mat{mk}t}\Vec{q}(0)+\int\limits_{0}^{t}\d t'\,
\e^{-\Mat{mk}(t-t')}\Mat{m}\tilde{\Vec{f}}(t') \, ,
\end{equation}
where the matrix exponential is defined, as usual, by its Taylor
expansion $\e^{-\Mat{mk}t}=\sum_{s=0}^{\infty}(-\Mat{mk}t)^{s}/s!$
Since we are only interested in the fluctuating part of $\Vec{q}(t)$,
we omit the deterministic relaxation $\e^{-\Mat{mk}t}\Vec{q}(0)$
due to an initial displacement by choosing $\Vec{q}(0)=\Vec{0}$.
In calculating the correlation matrix, we employ the second part of
Eq.~(\ref{eq:white_noise}) and consider the long-time limit 
($t\gg\lambda_{n}^{-1}$), where the system has relaxed to thermal 
equilibrium. We finally obtain 
\begin{equation}
\label{eq:correl-matrix}
\langle\Vec{q}(t + \tau)\otimes\Vec{q}(t)\rangle =
\kB T\,\e^{-\Mat{mk} \tau }\Mat{k}^{-1} \enspace, \enspace \tau \ge 0
\enspace.
\end{equation}
Note that the correlation matrix is symmetric, i.e., 
$\langle\Vec{q}(t_{1})\otimes\Vec{q}(t_{2})\rangle
=\langle\Vec{q}(t_{2})\otimes\Vec{q}(t_{1})\rangle$.

To express the correlation functions explicitly in coordinates, we
introduce the dual eigenvectors $\Vec{b}_{n}$ via
$(\Mat{mk})\T\Vec{b}_{n}=\lambda_{n}\Vec{b}_{n}$. Together with the
eigenvectors $\Vec{a}_{n}$, they fulfill the
orthonormality and completeness relations, 
$\Vec{a}_{m}\cdot\Vec{b}_{n}=\delta_{mn}$ and
$\sum_{n}\Vec{a}_{n}\otimes\Vec{b}_{n}=\Mat{1}$. Representing the
matrix exponential by its spectral decomposition, $\e^{-\Mat{mk}t}
=\sum_{n}\e^{-\lambda_{n}t}\Vec{a}_{n}\otimes\Vec{b}_{n}$, we obtain
from Eq.~(\ref{eq:correl-matrix})
\begin{equation}
\langle q_{m}(t+\tau)\,q_{n}(t)\rangle=\frac{\kB T}{k_{nn}}\,
\sum\limits_{l}\e^{-\lambda_{l}\tau}a_{lm}b_{ln} \,\, ,
\end{equation}
where $a_{lm}$ ($b_{ln}$) is the $m$th ($n$th) component of the
vector $\Vec{a}_{l}$ ($\Vec{b}_{l}$) and $k_{nn}$ the $n$-th element
of the diagonal matrix $\Mat{k}$
($k_{nn}=\kt,\kt,\kr,\kr$). Normalizing the correlation function to
the square roots of the mean square displacements $\langle
q_{n}^{2}\rangle=\kB T/k_{nn}$, we finally arrive at
\begin{equation}
\label{eq:norm_correl}
\frac{\langle q_{m}(t+\tau)\,q_{n}(t)\rangle}
{\sqrt{\langle q_{m}^{2}\rangle\langle q_{n}^{2}\rangle}}
=\sqrt{\frac{k_{mm}}{k_{nn}}}\,
\sum\limits_{l}\e^{-\lambda_{l}\tau}a_{lm}b_{ln} \,\, .
\end{equation}

\subsection{Longitudinal motions}
\label{subsec:correl_long}

Longitudinal motions are characterized by the eigenvectors and eigenvalues 
in Eq.~(\ref{eq:evec/eval_long}).
Translations and rotations are decoupled. So the respective auto- ($+$) and
cross- ($-$) correlations for rotational motions are given by
$[\exp(-\kr\murrpara_{+}\tau) \pm \exp(-\kr\murrpara_{-}\tau)]/2$, where
$\murrpara_{\pm}=\murrpara_{11}\pm\murrpara_{12}$. For translations, the
correlation functions are the same, however the relaxation rates are 
replaced by $\kt\muttpara_{\pm}$ \cite{mei99,bar01,hen01}.

\begin{figure}[b]
\centering
\includegraphics[width=\columnwidth]{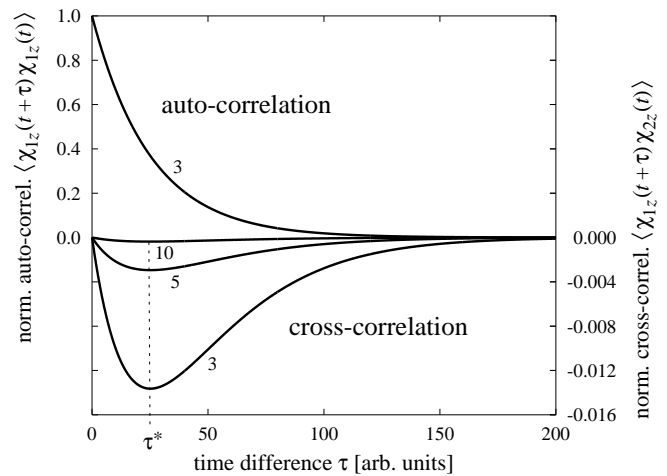}
\caption{Correlation functions for longitudinal rotational motions. 
The curve labels indicate the dimensionless center-to-center 
distance $\rho=r/a$. The functions are normalized to the mean square 
angular displacement $\langle \chi_{iz}^{2}\rangle=\kB T/\kr$.
The mobilities were calculated using the numerical library
{\sc hydrolib} \cite{hin95}.}
\label{fig:correl_long}
\end{figure}

The correlations for translational motions have already
been measured experimentally and compared with theoretical predictions
\cite{mei99,bar01,hen01}. These works reveal a good agreement between 
theory and experiment. The correlation functions for rotational motions 
show the same qualitative behavior (Fig.~\ref{fig:correl_long}), 
however with some quantitative differences.

First of all, we realize that the strength of the cross correlations
decreases with increasing particle separation. This is explained by
the decrease of the cross-mobility $\murrpara_{12}$, which is due to the 
spatial decay of the flow fields created by the particle rotation.
Compared to translational motions, this decrease is more pronounced since
rotational flow fields decay stronger than their translational 
counterparts.

Secondly, while the auto-correlation reveals the typical monotonous
decay, the cross-correlation function features an interesting
behavior. It is negative (denoted as ``anti-correlation'' in
Ref.~\cite{mei99}), since the relative modes
decay more slowly than the collective ones, as already mentioned in
Sec.~\ref{subsec:long_modes}. Therefore, the negative term in the
cross-correlation function in the first paragraph dominates.

The most interesting feature of the cross-correlation
function is that it exhibits a ``memory effect.'' It vanishes at
$\tau=0$, in contrast to what one would initially expect for the
instantaneous hydrodynamic forces in Stokesian dynamics, and then
shows a distinct time-delayed extremum (located at $\tau^{*}$ in
Fig.~\ref{fig:correl_long}). This behavior can be understood as
follows. The motion of particle~1 creates a fluid flow which
instantaneously reaches particle~2. However, due to the trap, particle~2
can only ``react'' in a finite time. Thus, the correlation evolves on a
characteristic time scale which is related to the relaxation times and
thereby to the trap stiffness. Since the ``memory'' is ``stored'' in the 
trap, it cannot last longer than the typical relaxation time. Hence, 
the correlation decays to zero for times larger than $\tau^{*}$.

For sufficiently large particle distances, the self-mobility
$\murrpara_{11}$ is much bigger than the cross-mobility
$\murrpara_{12}$. Then, to leading order, the characteristic time
scale is given by the single-particle rotational relaxation time 
$\tau^{*}=(\kr\mur)^{-1}$, in analogy to
the translational case \cite{mei99,bar01,hen01}.
Indeed, according to Fig.~\ref{fig:correl_long}, $\tau^{*}$ depends only 
very weakly on the particle distance. Nevertheless, the 
cross-mobility $\murrpara_{12}$ is sufficiently large to separate the 
two time scales $(\kr\murrpara_{+})^{-1}$ and $(\kr\murrpara_{-})^{-1}$, 
which is the origin of the ``anti-correlation''.

\subsection{Transversal motions}
\label{subsec:correl_trans}

The transversal eigenmodes are characterized by
Eqs.~(\ref{eq:evec/eval_trans_12}) and
(\ref{eq:evec/eval_trans_34}). Since the two types of eigenvectors 
are orthogonal to each other, i.e.,
$\Vec{a}_{1/2}\cdot\Vec{a}_{3/4}=0$, the 
dual vectors $\Vec{b}_{1/2}$ are linear combinations of $\Vec{a}_{1}$
and $\Vec{a}_{2}$ only; the equivalent holds for $\Vec{b}_{3/4}$.
One finds
\begin{equation}
\Vec{b}_{n}=
\frac{
|\Vec{a}_{\bar{n}}|^{2}\Vec{a}_{n}
-(\Vec{a}_{n}\cdot\Vec{a}_{\bar{n}})\Vec{a}_{\bar{n}}
}{
|\Vec{a}_{n}|^{2}|\Vec{a}_{\bar{n}}|^{2}
-(\Vec{a}_{n}\cdot\Vec{a}_{\bar{n}})^{2}
}
\,\, ,
\end{equation}
where the index combinations are $(n|\bar{n})=(1|2)$, $(2|1)$,
$(3|4)$, $(4|3)$.
The correlation functions for the coordinates $q_{1}=r_{1x}$,
$q_{2}=r_{2x}$, $q_{3}=\chi_{1y}$, and $q_{4}=\chi_{2y}$
are obtained from Eq.~(\ref{eq:norm_correl}). Now, they are linear
combinations of four exponential decays. 

The auto-correlation functions $\langle
r_{1x}(t+\tau)r_{1x}(t)\rangle$ and $\langle
\chi_{1y}(t+\tau)\chi_{1y}(t)\rangle$ show qualitatively 
the same monotonous decay as in the case of longitudinal fluctuations
(see Fig.~\ref{fig:correl_long}). All the other correlation functions
exhibit the features of the longitudinal cross-correlation discussed in
Sec.~\ref{subsec:correl_long} (except for the sign).

\begin{figure}[t]
\centering
\includegraphics[width=\columnwidth]{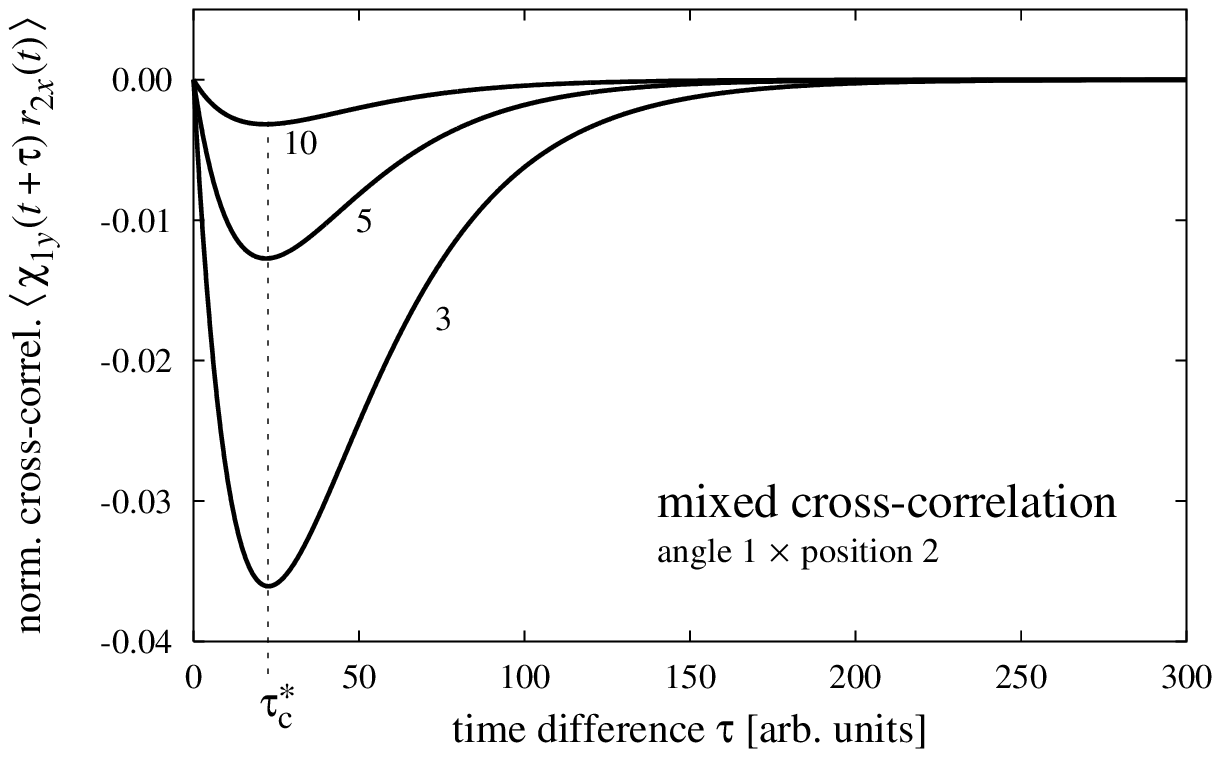} \\
\includegraphics[width=\columnwidth]{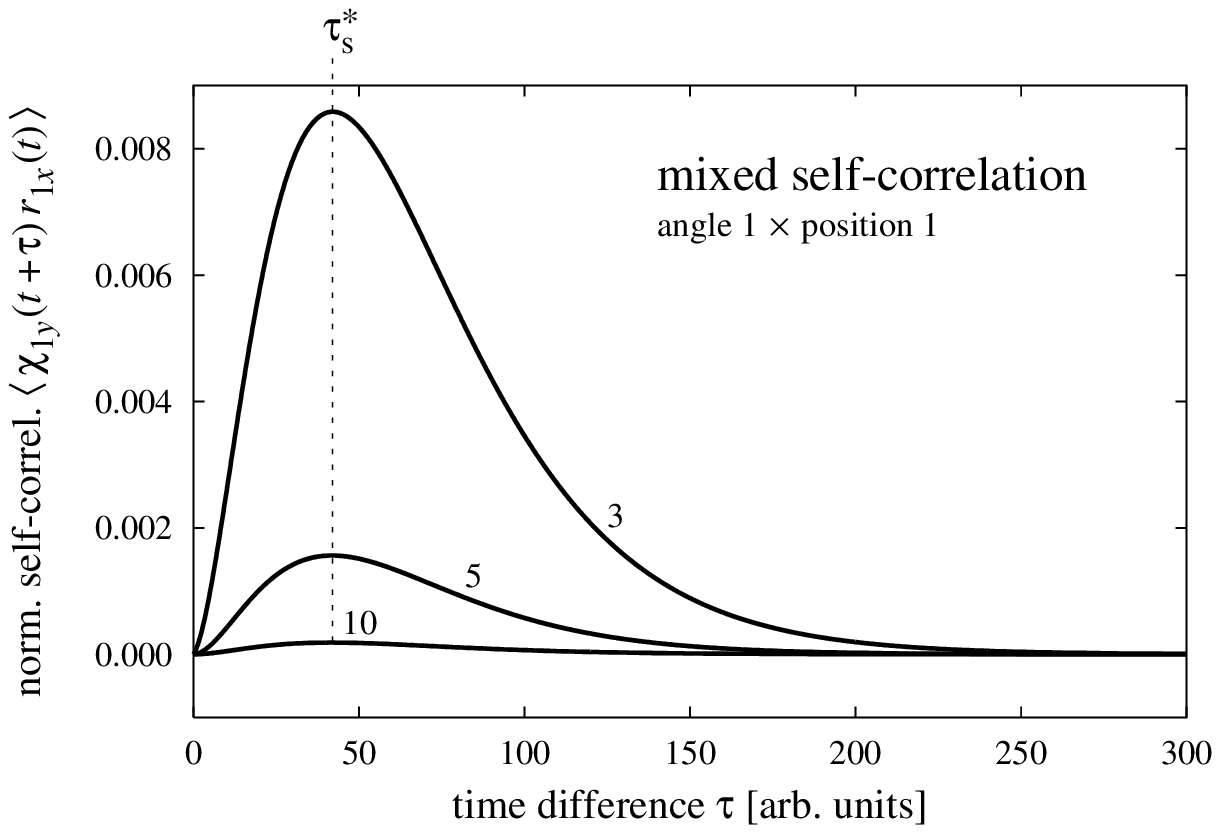}
\caption{Correlation functions for transversal motions. The curves
shown are the mixed cross- and self-correlation for angle and position.
The curve labels indicate the dimensionless center-to-center distance
$\rho=r/a$. The trap constants are chosen as $\kt=\kr$. The functions
are normalized to the square roots of the 
mean square displacements $\langle\chi_{iy}^{2}\rangle=\kB T/\kr$ and 
$\langle r_{ix}^{2}\rangle=\kB T/\kt$ [see Eq.~(\ref{eq:norm_correl})].}
\label{fig:correl_trans}
\end{figure}

As representative examples, the correlation functions for rotations
about the $y$-axis and translations along the $x$-direction,
$\langle\chi_{1y}(t+\tau)r_{1x}(t)\rangle$ and
$\langle\chi_{1y}(t+\tau)r_{2x}(t)\rangle$, are plotted in
Fig.~\ref{fig:correl_trans}. In the following, we will refer to them
briefly as mixed self- and cross-correlation, describing the self- and
cross-coupling of rotation and translation, respectively.

Both correlation functions show a distinct time-delayed extremum,
denoted by $\tau_{\text{s}}^{*}$ and $\tau_{\text{c}}^{*}$ in
Fig.~\ref{fig:correl_trans}. The mixed cross-correlation is interpreted
in the same fashion as the longitudinal cross-correlation in the
preceeding section.

The most striking feature is revealed by the mixed
self-correlation. For a single sphere, translation and rotation are
not coupled \cite{hes68}, i.e., the mixed (self-)correlation
vanishes. However, in the two-sphere system, there exists a mixed 
self-correlation, as shown in Fig.~\ref{fig:correl_trans}, that is
mediated by the neighboring particle. This correlation is weaker
than the mixed cross-correlation since the flow field created by
particle~1 has to be reflected by particle~2. As discussed before,
particle~2 ``reacts'' with a finite delay due to the trap
stiffness. Then, in addition, particle~1 also has a finite ``reaction
time.'' Hence, we always expect
$\tau_{\text{s}}^{*}>\tau_{\text{c}}^{*}$.

\begin{figure}[t]
\centering
\includegraphics[width=\columnwidth]{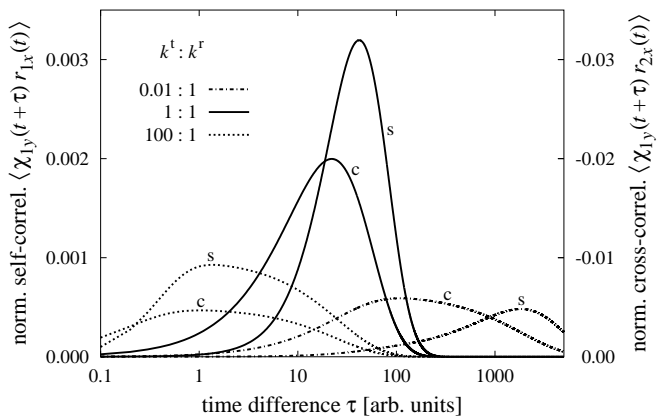}
\caption{Mixed self- (label s) and cross-correlation (label c) for
various trap constant ratios $\kt:\kr$. The force constant $\kt$ was
varied ($\kt=0.01$, $1$, and $100$), while the torque constant was kept
fixed ($\kr=1$). The qualitative behavior of the functions (discussion
see text) is the same for different particle distances (here: $\rho=4$).}
\label{fig:correl_trap-depend}
\end{figure}

We also studied the influence of the trap stiffnesses $\kt$ and
$\kr$ on the correlation functions. As illustrated in
Fig.~\ref{fig:correl_trap-depend}, the difference in the delay times
$\tau_{\text{s}}^{*}$ and $\tau_{\text{c}}^{*}$ decreases with
increasing ratio $\kt/\kr$. Furthermore, we observe that
the correlations are strongest for $\kt\approx\kr$.

Note that the mixed self-correlations cannot be treated within the
Rotne-Prager approximation. They constitute an additional effect
of higher order, as mentioned in Sec.~\ref{subsec:rotne-prager}. 

\section{Conclusions}

In this paper, we have presented the complete solution for the 
hydrodynamically coupled translational and rotational motions of two 
colloidal spheres that are harmonically trapped with respect to both 
their positions and orientations. 

Based on pure symmetry arguments and
without relying on explicit values of the mobilities, we have
determined all the 12 collective eigenmodes and qualitatively
discussed their relaxation times. Whereas the properties of the 
longitudinal modes are reminiscent to a system with pure
translational degrees of freedom, the transversal modes exhibit
a characteristic coupling of translation and rotation.

In a detailed Langevin-type analysis, we have been able to derive the
full set of correlation functions characterizing the Brownian motion
of the particles in the optical traps. The analysis relies on the
eigenvalue problem of nonsymmetric matrices. Explicit examples for the
correlation functions at different particle separations have been
calculated based on mobilities which we obtained from the 
numerical library {\sc hydrolib} \cite{hin95}. 

The longitudinal 
fluctuations exhibit the features already mentioned in
Refs.~\cite{mei99,bar01,hen01}, namely a ``memory effect'' in the
cross-correlations. The transversal fluctuations are governed by the
coupling of translations and rotations.
As the most striking feature, this coupling is also visible in the
self-correlations which are governed by a second delay time in addition
to the one observed in the cross-correlations. The self-coupling of 
translation and rotation for one sphere has to be mediated by a
second sphere. It is not included in the Rotne-Prager
approximation and therefore introduces an additional effect of
higher order.

Correlation functions involving the rotational degrees
of freedom are weaker compared to pure translational correlations since
the flow field of a single rotating sphere decays as $1/r^{2}$, compared to
translating particles where the decay is $1/r$. However, the strength
of the correlations increases for decreasing particle separation
which is more pronounced whenever rotations are involved.
Furthermore, at sufficiently small
distances, lubrication theory becomes important \cite{kim85}, and the
system introduced in this paper may help to check its predictions.

Our work stresses the rotational degree of freedom and its influence
on hydrodynamic interactions. We hope to stimulate experimental
investigations of the correlation functions presented in this
paper. 

In the Introduction, we have already mentioned the field of
microrheology. An extension of our work to viscoelastic media based on
the theoretical approach presented in Ref.~\cite{lev00} seems
appealing. The rotational motion of the probe particle results in an
interesting deformational mode since it introduces some torsion in
the surrounding medium. It therefore can yield additional information
about the viscoelastic properties.

\appendix

\begin{acknowledgments}
We would like to thank Paul Bartlett, Thomas Gisler, Georg Maret, and
Stephen Martin for fruitful discussions.
This work was supported by the Deutsche Forschungsgemeinschaft through
the Sonderforschungsbereich Transregio~6 ``Physics of colloidal
dispersions in external fields.''
\end{acknowledgments}

\end{document}